\documentclass[letter]{aa} 

\usepackage{graphicx}
\usepackage{txfonts}
\begin{document}

   \title{Quasars as high-redshift standard candles}

   \subtitle{}

   \author{A. Sacchi\inst{1}\thanks{\email{andrea.sacchi@unifi.it}}, G. Risaliti\inst{1,2}, M. Signorini\inst{1,2}, E. Lusso\inst{1,2}, E. Nardini\inst{2}, G. Bargiacchi\inst{3,4}, S. Bisogni\inst{5}, F. Civano\inst{6}, M. Elvis\inst{6}, G. Fabbiano\inst{6}, R. Gilli\inst{7}, B. Trefoloni\inst{1,2} \and C. Vignali\inst{8,7}}

   \institute{Dipartimento di Fisica e Astronomia, Universit\'a di Firenze, via G. Sansone 1, 50019 Sesto Fiorentino, Firenze, Italy
         \and
             INAF - Osservatorio Astrofisico di Arcetri, Largo Enrico Fermi 5, I-50125 Firenze, Italy
         \and
             Scuola Superiore Meridionale, Largo S. Marcellino 10, 80138 Napoli, Italy
         \and
             Istituto Nazionale di Fisica Nucleare (INFN), Sez. di Napoli, Complesso Univ. Monte S. Angelo, Via Cinthia 9, 80126, Napoli, Italy
         \and
             INAF – Istituto di Astrofisica Spaziale e Fisica Cosmica Milano, Via Corti 12, 20133 Milano, Italy
         \and
             Harvard-Smithsonian Center for Astrophysics, 60 Garden Street, Cambridge, MA 02138, USA
         \and
             INAF – Osservatorio di Astrofisica e Scienza dello Spazio di Bologna, via Gobetti 93/3, I-40129 Bologna, Italy
         \and
             Dipartimento di Fisica e Astronomia "Augusto Righi", Universit\'a degli Studi di Bologna, via Gobetti 93/2, I-40129 Bologna, Italy
             }
  \authorrunning{A. Sacchi, G. Risaliti, M. Signorini, E. Lusso, E. Nardini}

   \date{Received xxx; accepted xxx}

 
  \abstract{The non-linear relation between the X-ray and ultraviolet (UV) luminosity in quasars has been used to derive quasar distances and to build a Hubble diagram at redshifts up to $z$\,$\sim$\,7. This cosmological application is based on the assumption of independence of the relation on redshift and luminosity. We want to test the reliability of this hypothesis by studying the spectroscopic properties of high-redshift quasars in the X-ray and UV bands. We performed a one-by-one analysis of a sample of 130 quasars at $z$\,$>$\,2.5 with high-quality X-ray and UV spectroscopic observations. We found that not only the X-ray to UV correlation still holds at these redshifts, but its intrinsic dispersion is as low as 0.12~dex (previous works reached 0.20--0.22~dex). For a sample of quasars at $z$\,$\sim$\,3 with particularly high-quality observations the dispersion further drops to 0.09~dex, a value entirely accountable for by intrinsic variability and source geometry effects. The composite spectra of these quasars, in both the X-rays and the UV, do not show any difference with respect to the average spectra of quasars at lower redshifts. The absence of any spectral difference between high- and low-$z$ quasars and the tightness of the X-ray to UV relation suggests that no evolutionary effects are present in the relation. Therefore, it can be safely employed to derive quasar distances. Under this assumption, we obtain a measurement of the luminosity distance at $z$\,$\sim$\,3 with 15\% uncertainty, and in a $4\sigma$ tension with the concordance model.}

   \keywords{Quasars --
                cosmology --
                High energy astrophysics
               }

   \maketitle

\section{Introduction}

   Quasars are the most luminous persistent sources in the Universe, and have been identified up to redshifts $z$\,$>$\,7 \citep{mor11,ban17,wan21}. At present, more than half a million of them have been spectroscopically identified by the Sloan Digital Sky Survey \citep{lyk20}. For about 15,000 such sources, an X-ray (mostly serendipitous) observation is also available in the {\em XMM-Newton} and/or {\em Chandra} archives. Starting from this sample, selected sources have been used in the past few years to investigate the non-linear relation linking the X-ray and ultraviolet (UV) luminosities in quasars, usually expressed as $\log L_\textup{X} = \gamma\log L_\textup{UV}+\beta$, where $L_\textup{X}$ and $L_\textup{UV}$ are the monochromatic luminosities at the rest-frame 2 keV and 2500 \AA\,\,and $\gamma\approx0.6$ \citep{tan79}. The non linearity of the relation allows us to derive quasar distances from their X-ray and UV flux densities, and to build a Hubble diagram of quasars up to $z$\,$\sim$\,7 \citep{ris19,lus20}. In the disc-corona framework, the UV emission of quasars comes from  an accretion disc, where the gravitational energy of the accreting gas is converted into radiation \citep{sha73}, while the X-ray emission comes from the inverse Compton scattering of a fraction of the UV photons by a ``corona'' of hot plasma \citep{haa93}. Since the X-ray corona would cool down rapidly as a consequence of the inverse-Compton process, while the X-ray emission is persistent, it is obvious that a physical process continuously transferring energy from the disc to the corona must be present in quasars. 
   Even if no complete physical model of this process is at present available, the observed non-linear $L_\textup{X}-L_\textup{UV}$ relation must be a consequence of such a mechanism. 
   The key requirement regarding the possible cosmological application of the relation is therefore that it holds at all redshifts and luminosities with no evolution. 
   We note that the redshift-independence of the slope was clearly observed by independent groups well before this relation was employed for cosmological studies and with quasar samples selected in various ways, e.g., \citet{vignali03} (see their Section 4.4), \citet{steffen06} (see their Fig. 9 and Section 3.5), \citet{just07} (see their Fig. 9 and related discussion), \citet{green09} (see their Section 7).
   
   The state of the art in the study of the X-ray to UV relation in quasars is presented in \citet{lus20} and \citet{bis21}, hereafter referred to as L20 and B21, respectively. The two samples analysed in these works mainly consist of sources with serendipitous {\em XMM-Newton} or {\em Chandra} X-ray observations and SDSS optical/UV spectra. The flux densities were estimated from the photometric data points in both the optical/UV and X-ray bands in the former work, while in the latter the X-ray flux densities were obtained through an automated spectroscopic analysis. In both works the X-ray to UV luminosity relation is analysed by splitting the sample in narrow redshift intervals. Since the spread in distance within a given bin is small with respect to the observed dispersion, it is possible to reproduce the relation using fluxes as proxies of luminosities. This allows us to investigate the possible redshift evolution of the slope of the relation in a cosmology-independent way. 
   The main results of these studies are that {\it a)} the relation holds over the whole $z$\,$\sim$\,0.5--7.5 redshift range, with an average intrinsic dispersion of 0.20--0.22~dex; {\it b)} the measured values of $\gamma$ do not show any hint of evolution, and are consistent with a constant value of $\sim$0.6; {\it c)} the Hubble diagram of quasars derived using the relation is in perfect agreement with that of supernovae Ia;  
   {\it d)} the  Hubble diagram of quasars in the whole $z$\,$=$\,0.5--7.5 range is in a $>$\,4$\sigma$ tension with the standard flat $\Lambda$CDM model \citep{ris19,lus20,bar21}. 
   
   The results on the Hubble diagram critically depend on the assumption of non-evolution of the relation. On the one hand, the constancy of the slope and the agreement with supernovae (and, hence, with the flat $\Lambda$CDM model) up to $z$\,$\sim$\,1.5 suggests a non-evolution. On the other hand, it has been shown by assuming several different cosmological models that a tension exists between the parameters of the relation at low and high redshift \citep{kr2021}. 
   This result can be interpreted as due to either a redshift evolution of the relation, if the models tested by \citet{kr2021} include the ``true'' one, or a different cosmological model, if the relation does not evolve with redshift.
   
   In order to clarify this point it is fundamental to further investigate the relation with the main goals of (1) ruling out possible redshift-related systematic effects in the determination of distances, and (2) reducing the observed dispersion, in order to improve its effectiveness as a cosmological probe. 
   We will show that these two goals are closely intertwined: in fact, reducing the observed dispersion is an effective way to rule out major systematic effects.

   Here we present the most detailed study to date of the X-ray to UV relation at high redshift, considering a sample of 130 quasars at 2.5\,$<$\,$z$\,$<$\,4.5. The two aims described above were pursued in the most direct way: we performed a complete UV and X-ray spectroscopic analysis of all the sources in the sample, in order to improve the accuracy of flux measurements. The sample covers an optimal redshift interval to test possible deviations of the Hubble diagram from the standard cosmological model (see also \citealt{lus19}), and its size balances the need for a high enough statistics and the feasibility of a complete spectroscopic analysis.

\section{Data selection and analysis}

   Full details of the selection and the data analysis are reported in L20 and B21. Here we briefly summarize our method. The 130 quasar sample under analysis is drawn from three main catalogues:\\
   \noindent \underline{\smash{\it SDSS-XMM.}} The bulk of our sample is composed by 81 quasars from the cross-match between the SDSS DR16 Quasar catalog \citep{fle21} and the {\em XMM-Newton} Serendipitous Source Catalogue Data Release 10 \citep{web20}. Within this group, 15 quasars have pointed {\em XMM-Newton} observations and compose the {\it XMM-pointed} subsample. These sources have been extensively described in a dedicated paper \citep{nar19}. 
   The remainder of the {\it SDSS-XMM} subsample hence amounts to 66 sources. The sample selection follows the procedure described in L20, and consists of the following filters:\\ 
   1) Removal of all the known Broad Absorption Line (BAL) and radio-loud objects as for both classes of objects the intrinsic emission of both disc and corona is contaminated by other physical phenomena (e.g. gas absorption, jets), dimming or enhancing in particular their X-ray luminosities \citep{bra00,wil87}.\\
   2) Removal of all the sources with an X-ray spectroscopic photon index $\Gamma-\Delta\Gamma<1.7$ and $\Gamma+\Delta\Gamma>2.5$. This ensures that no objects with photoelectric absorption (which would cause a flattening of the X-ray spectra) or calibration issues (which could produce a spectral steepening) are present.\footnote{Intrinsically super-soft spectra are extremely rare in quasars, and would require anomalous coronal properties (cf. equation A1 in \citealt{zdz96}).} \\
   3) Removal of all the sources with optical/UV colours outside the region defining ``normal'', blue quasars without significant dust absorption. The threshold, corresponding to $E(B-V)$\,$\simeq$\,0.1, is the same as in L20.\\ 
   4) Removal of all the sources with an expected X-ray flux close to the detection limit of its observation. Sources with an intrinsic flux close to the observable limit will be detected, because of their variability, only if they are caught in a positive fluctuation. As less luminous quasars show more significant variability, this would affect the slope of the $L_\textup{X}$--$L_\textup{UV}$ relation. Again, the threshold is the one adopted in L20.\\
   \noindent \underline{\smash{\it SDSS-Chandra.}} Sixteen quasars come from the cross-match of the Sloan Digital Sky Survey quasar catalog Data Release 14 \citep{par18} and the Chandra Source Catalog 2.0 \citep{eva10}. The selection is analogous to the one described above for {\em XMM-Newton} sources, and is fully described in B21.\\
   \noindent \underline{\smash{\it COSMOS.}} Thirty-three quasars are extracted from the {\em Chandra}-COSMOS Legacy Survey \citep{civ16}. The sample selection and calculation of UV and X-rays fluxes are fully described in B21.\\

\subsection{Flux and luminosity measurements}

   The sources from the {\it COSMOS} and {\it XMM-pointed} subsamples have been fully described in published works \citep{mar16,nar19,bis21}. We did not repeat the analysis of either the X-ray spectrum or the optical one, and we adopted the reported values. \\
   For all the other 82 sources (66+16 from the {\it SDSS-XMM} and {\it SDSS-Chandra} subsamples), we performed a complete analysis of the SDSS optical (rest-frame UV) spectra and of the {\em XMM-Newton} or {\em Chandra} X-ray spectra.
   
   For the X-ray measurements, we followed the standard procedure described in the user manuals of the two observatories, and we analysed the spectra using the XSPEC code \citep{arn96}. We assumed a power-law model with Galactic photoelectric absorption. We show the spectra and best fit models for a subsample of 12 sources (see below) that were not previously published in Appendix~\ref{X-ray and optical/UV spectra}.
   
   The optical SDSS spectra have been analysed with the IDL package QSFit \citep{cal17}. Using this software we modeled the features of each quasar spectrum, by assuming the continuum to be a power law, the emission lines to have Gaussian shapes and the host galaxies to be ellipticals (the interested reader can find more details in \citealt{cal17}). The strength of this code relies in the fact that the spectral properties are all fitted simultaneously, which avoids the fitted parameters to depend on local features of the spectrum. For each object, we derived the continuum and emission-line properties. The rest-frame 2500 \AA~wavelength is out of the SDSS range because of the redshift of these objects. Therefore, to derive the monochromatic flux at 2500 \AA, we extrapolated  from the continuum shape, which is assumed to be a single power law whose slope is one of the fitted parameters. We show the resulting plots of the described fit procedure for the same 12 sources mentioned above in  Appendix~\ref{X-ray and optical/UV spectra}. 
   The luminosity values, when reported, are computed by assuming a standard flat $\Lambda$CDM cosmology with $\Omega_\textup{M}=0.3$ and $H_0=70$ km s$^{-1}$ Mpc$^{-1}$.

\subsection{X-ray cross-calibration}
   As the X-ray fluxes of both the {\it SDSS-XMM} and {\it SDSS-Chandra} subsamples have been obtained by spectroscopic analysis, no cross-calibration is needed in order to compare them. In principle, the {\it COSMOS} subsample would instead require such a calibration, as we derived the 2 keV flux densities from the soft- and hard-band fluxes available in the published catalog \citep{mar16}. Yet, we decided not to apply it. This choice is based on two main reasons. On the one hand, we tried to estimate the parameters of cross-calibration by fitting the {\it COSMOS} X-ray fluxes vs.~the {\it XMM-Newton} ones for the sources appearing in both parent samples with a linear model, $\log F_\textup{XMM}+31.5=a\,(\log F_\textup{COSMOS}+31.5)+b$. Over $\approx$\,1500 observations, we retrieved values of $a=0.89$ and $b=0.10$, which do not affect significantly our estimates of the slope, dispersion, and offset of the relation. On the other hand, the factors to be taken into account when attempting a precise cross-calibration are too many (e.g., the evolution of the {\it Chandra} observatory effective area over the years, the off-axis angle and duration of each observation, the intrinsic shape and variability of the source spectra) to be reliably addressed with our sample statistics.

\subsection{Regression analysis} 
   The best fit of the $\log L_\textup{X}$--$\log L_\textup{UV}$ relation has been performed using the Python package {\scshape emcee} \citep{for13}, which is an implementation of Goodman \& Weare’s Affine Invariant Markov chain Monte Carlo (MCMC) Ensemble sampler. The UV and X-ray monochromatic luminosities were normalized to $10^{31.5}$ and $10^{27.5}$ erg s$^{-1}$ Hz$^{-1}$, respectively, prior to the regression fit. 
   For the full sample, we employed a sigma-clipping set to $2.7\sigma$, and this choice eliminated 7 outliers.

\section{Discussion}

\begin{figure}
   \centering
   \includegraphics[width=\hsize]{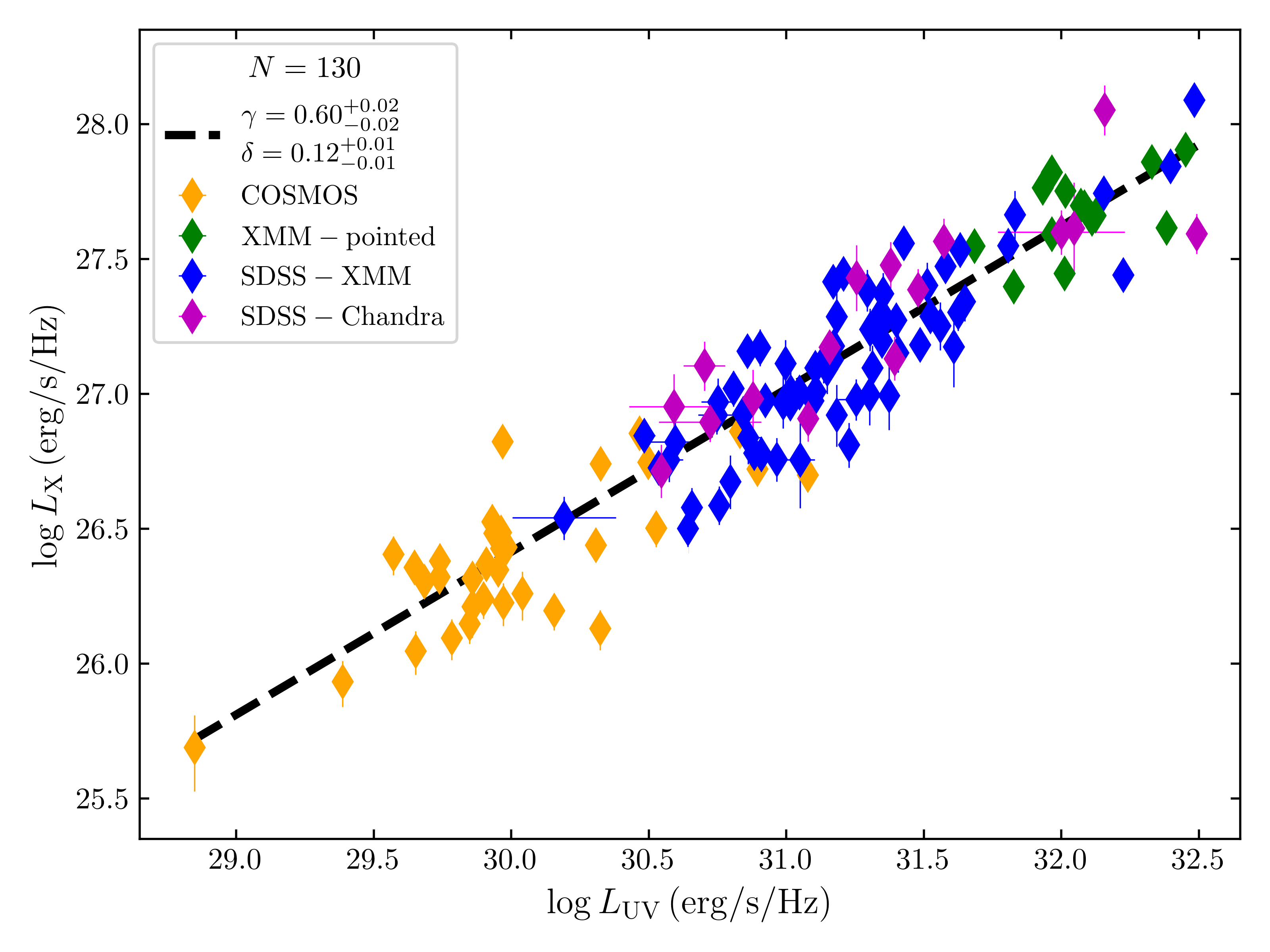}
      \caption{$\log L_\textup{X}$ vs.~$\log L_\textup{UV}$ for the full $z$\,$>$\,2.5 sample presented in this work. Colours refer to the different subsamples, as indicated in the legend. The $\gamma$ and $\delta$ symbols represent the slope and the intrinsic dispersion of the best fit relation, respectively.}
         \label{fig:full_sample}
   \end{figure}

  All the 130 quasars selected at $z$\,$>$\,2.5 have homogeneous UV and X-ray spectral properties, and high-quality X-ray observations. The latter requirement is key to determine the overall accuracy of the distance estimates, as for all the sources in the sample the rest-frame UV data provide higher-quality flux measurements than the X-ray ones.
  The results of the analysis are shown in Figure \ref{fig:full_sample}. The intrinsic dispersion $\delta$ of the $L_\textup{X}$--$L_\textup{UV}$ relation (i.e., the dispersion not accounted for by the statistical errors) is as low as 0.12~dex over more than three orders of magnitude in UV luminosity. This is a significant improvement with respect to the previous results ($\delta=0.21$, L20). The best-fit slope $\gamma=0.60\pm0.02$ is fully consistent with the values found at lower redshift (L20, B21), further confirming the stability of the relation with redshift.
  
  We note that the conversion of observed fluxes to luminosities requires a distance--redshift law. As a consequence, the relation as plotted in Figure~\ref{fig:full_sample} cannot be used to test cosmological models. However, this is not a critical issue for our purposes: here we only want to examine the relation, and the effects of using even very different cosmological models are almost entirely subsumed in the normalization parameter.

\begin{figure}
   \centering
   \includegraphics[width=\hsize]{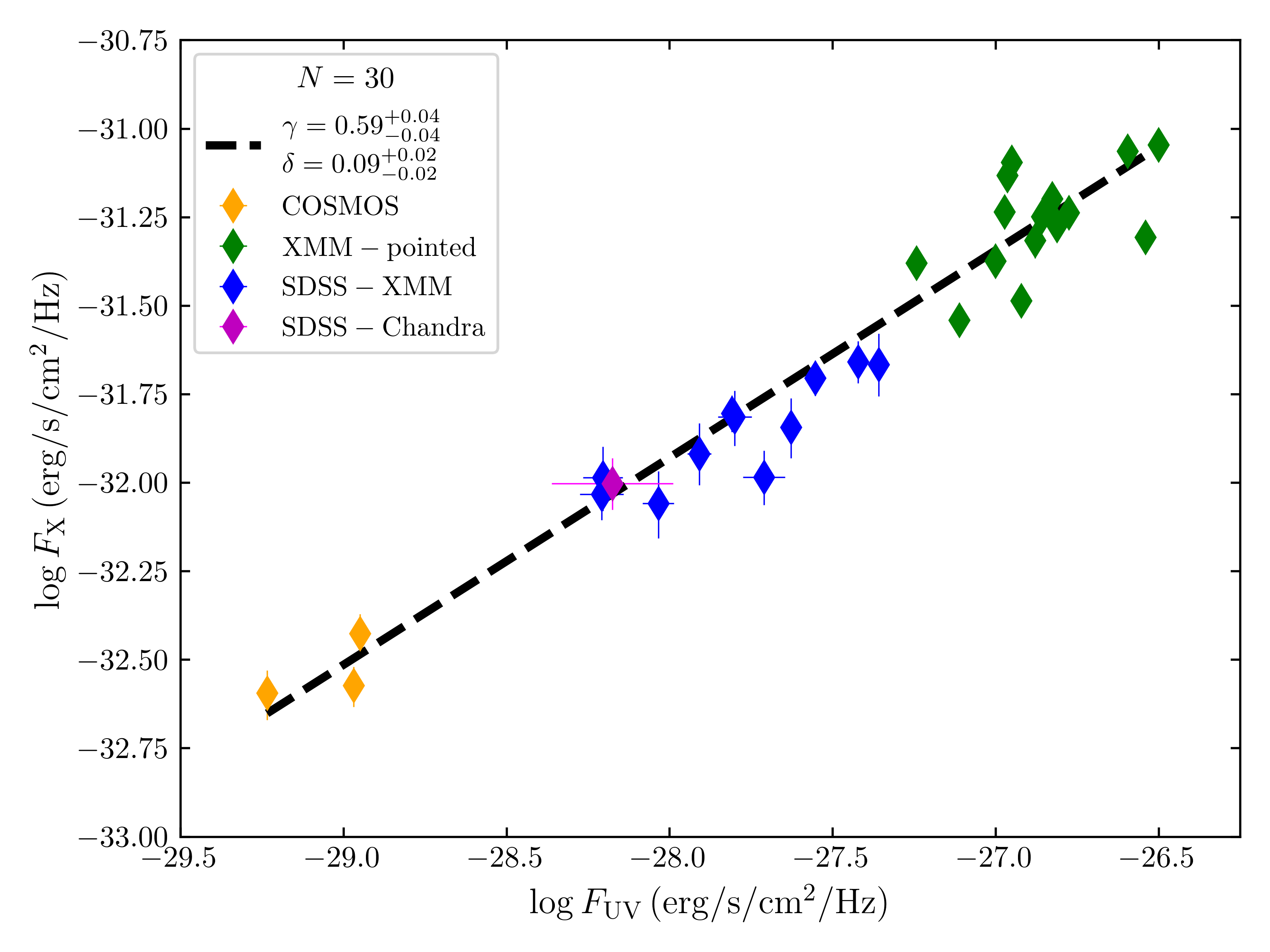}
      \caption{$\log F_\textup{X}$ vs.~$\log F_\textup{UV}$ for the $3.0$\,$<$\,$z$\,$<$\,3.3 ``golden'' sample. Colours refer to the different subsamples. The green points are the group of {\em XMM-Newton} pointed observations.}
         \label{fig:golden_bin}
   \end{figure}

   In order to obtain an even cleaner result, and to test cosmological models, we repeated our analysis for a ``golden'' subsample of 30 sources\footnote{We note that no source in the ``golden'' sample is removed by the sigma-clipping, as no outlier is present.} in the narrow redshift range $z=3.0$--3.3. This is the interval containing the group of 15 sources for which we have obtained high-quality {\em XMM-Newton} pointed observations \citep{nar19}, which compose the {\it XMM-pointed} subsample. 
   The X-ray to UV relation for this sample is shown in Figure \ref{fig:golden_bin}. Here the redshift interval is narrow enough to allow the analysis of the relation using the observed fluxes as proxies of luminosities. The intrinsic dispersion is now only 0.09~dex. Considering the unaccounted dispersion due to variability (of the order of at least 0.05~dex at these redshifts, \citealt{pao17}) and of the inclination of the disc (assuming that the X-ray emission is isotropic, while the UV emission is disc-like), we conclude that the relation has very little, if any, intrinsic dispersion: quasars are indeed standardizable candles.

\begin{figure}
   \centering
   \includegraphics[width=\hsize]{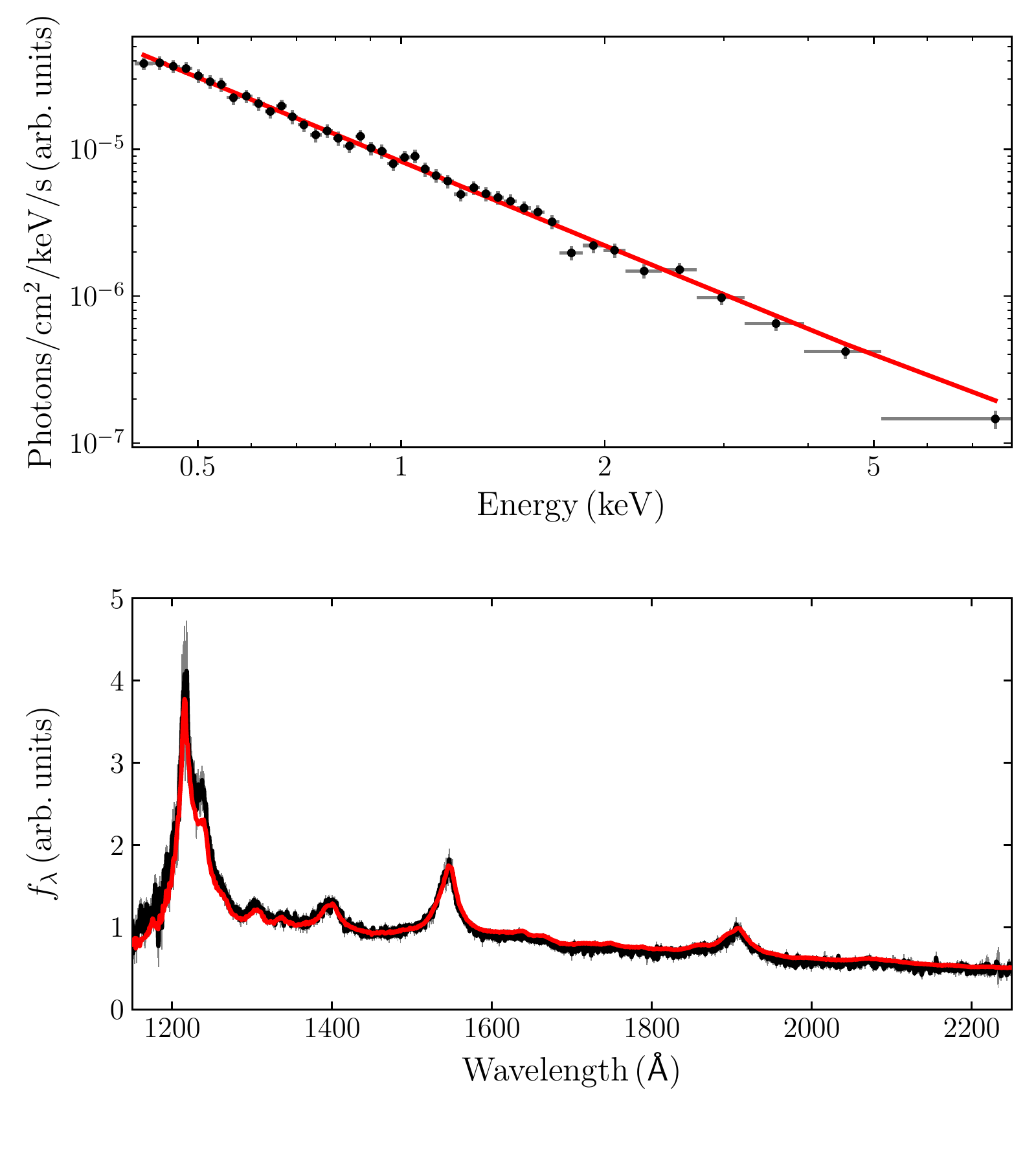}
      \caption{Average X-ray and UV spectral properties of the {\it SDSS-XMM} and {\it SDSS-Chandra} subsamples in the $z = 3.0$--3.3 redshift range. Upper panel: stacked X-ray emission (black data point); its power-law model photon index $\Gamma=1.89\pm0.02$ is perfectly compatible with the average value of $\Gamma=1.9$ of ``typical'' unobscured quasars \citep{ree00,ris09,sco11} shown by the solid red line. Lower panel: stacked SDSS spectrum (solid black line) compared with the average quasar spectrum (solid red line) from a large sample of SDSS quasars over a wide redshift and luminosity range \citep{van01}.}
         \label{fig:stacks}
   \end{figure}

   The complete X-ray and UV spectroscopic analysis for this sample allowed us to check for possible systematic effects in the spectra, a physical redshift evolution of their intrinsic properties and/or the consequence of dust/gas absorption. The results are shown in Figure \ref{fig:stacks}, where we overplot the rest-frame stacked X-ray and UV spectra of the {\it SDSS-XMM} and {\it SDSS-Chandra} quasars, normalized to their integral flux in the observed band (see \S~3 in \citealt{lus2015} for details on the stacking procedure). The stacked X-ray spectrum, with a slope $\Gamma=1.89\pm0.01$, shows no deviation from the $\Gamma$\,$\simeq$\,1.9 power-law model representing the average X-ray emission of quasars \citep{ree00,ris09,sco11}. Analogously in the UV, the absence of deviations amongst the individual spectra, and with respect to the average quasar spectrum of \citet{van01}, demonstrates that there is no systematic effect in the UV measurements. Since the latter composite spectrum is obtained from a sample of quasars in a wide redshift interval ($z=0.044$--4.789, with a median of $z\approx1.25$), we also demonstrate that there is no spectral evolution with redshift between local quasars (used to calibrate the relation) and those at $z$\,$\sim$\,3.
 
\begin{figure}
   \centering
   \includegraphics[width=\hsize]{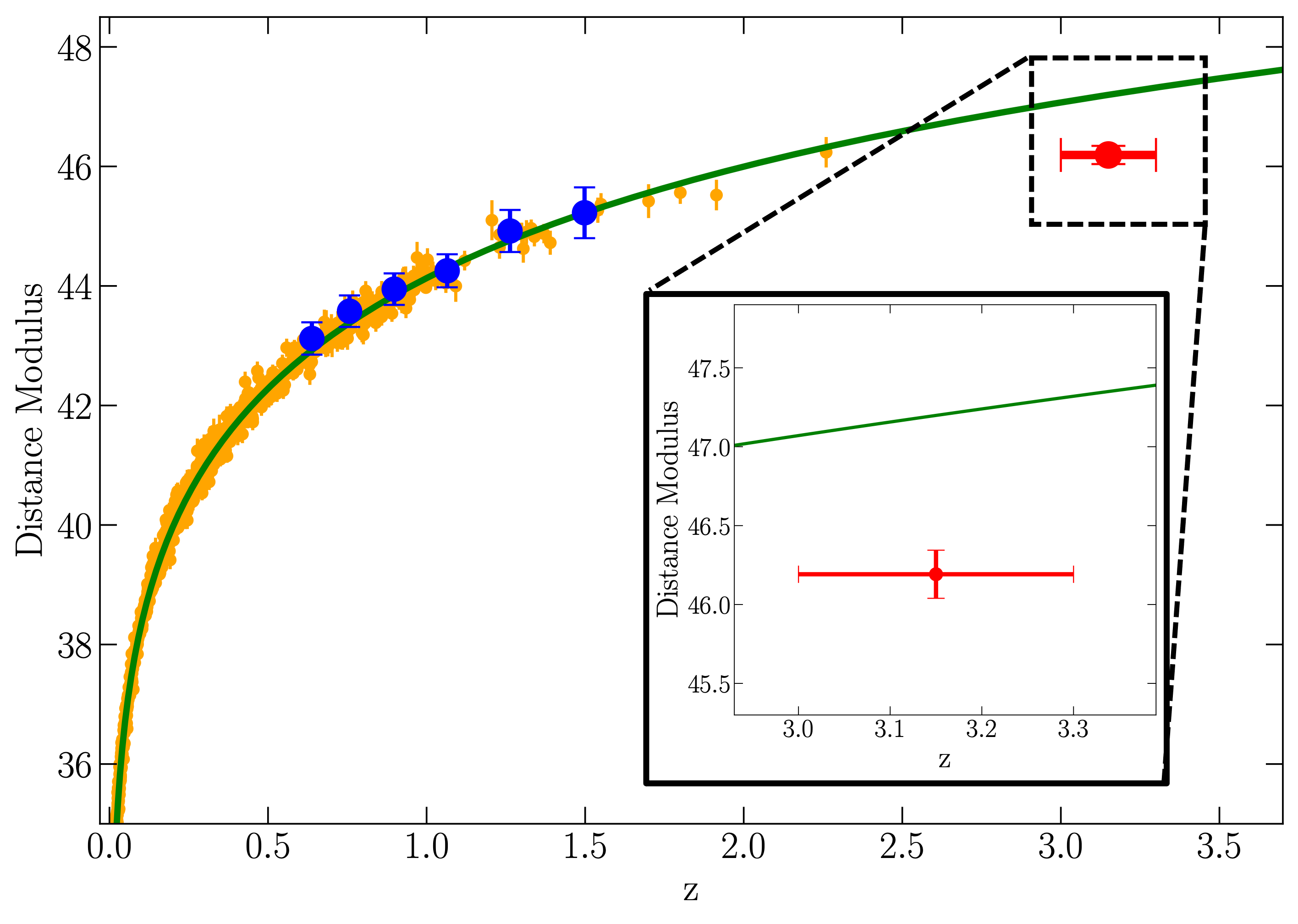}
      \caption{Hubble Diagram of {\em Pantheon} supernovae (orange points, \citealt{sco18}), quasars at redshifts $z=0.7$--1.3 (blue points), and quasars at redshifts $z=3.0$--3.3 (red point). The luminosity distances for quasars are calculated using the parameters $\gamma$ and $\beta$ as described in the text, i.e. assuming that these parameters do not change with redshift, and adopting the best-fit flat $\Lambda$CDM model for supernovae. Each quasar point represents the average for all the quasars in the corresponding redshift interval.}
         \label{fig:hubble}
   \end{figure}

   The final application of the work presented here is a direct test of the flat $\Lambda$CDM model, under the physical assumption of no redshift evolution of the normalization $\beta$ above $z=1.5$. We remind that the non-evolution of $\beta$ at $z<1.5$, and the non-evolution of the slope $\gamma$ at any redshift, are observationally proven by several independent groups. 
   The result in Figure \ref{fig:golden_bin} alone is not sufficient as a cosmological probe, due to the lack of an absolute calibration of the relation. This limit can be overcome by anchoring the intercept of the Hubble diagram built with quasars with other distance indicators at lower redshift (see \citealt{moresco2022} for a review), for example supernovae. We therefore built our Hubble diagram by including the latest sample of SNIa up to redshift $z\sim2$ (i.e. the {\em Pantheon} sample; \citealt{sco18}), averages of quasars at $z=0.7-1.3$, and the average for the ``golden'' sample (Figure \ref{fig:hubble}). For all the quasars in this diagram, the luminosity distances are calculated from the X-ray to UV flux-flux relation cross-calibrated with supernovae in the $z=0.7$--1.3 range. The distance modulus for the golden sample is then calibrated by fixing the parameters involved in the computation of the luminosity distance to the ones at low redshift (i.e. $\gamma=0.6$ and $\hat\beta=13.4$\footnote{The luminosity distance, $D_{\rm L}$ (in Mpc), is computed as  $\log D_{\rm L}=1/(2-2\gamma)(\gamma \log F_{\rm UV}-\log F_{\rm X}+\hat\beta)$, where $\hat\beta$ represents the intrinsic normalization of the distance modulus-redshift relation.}). This methodology implies that the best-fit parameters of the assumed cosmological model (i.e. flat $\Lambda$CDM with $\Omega_{\rm M}=0.3$) do not change with redshift.
   The observed tension between the average luminosity distance of the ``golden'' sample and the prediction of the flat $\Lambda$CDM model is $>$\,4$\sigma$.
   As a further test, we performed another fit of the Hubble diagram by using a fourth order polynomial function of the luminosity distance in $\log(1+z)$ as discussed in \citet{bar21}. We followed a similar approach as discussed above, by cross-calibrating the quasars at $z=0.7$--1.3 with SNIa. We then compared the distance modulus of the $z\simeq3$ sample to the resulting best-fit of the polynomial expansion when parameters are fixed to the best-fit ones of the low-redshift data. The observed tension is again confirmed at more than the $4\sigma$ statistical level.

\section{Conclusions}

   We presented a sample of 130 quasars at $z>2.5$ for which we performed a complete X-ray and UV spectral analysis. The sample was blindly selected following our well established method that excludes absorbed and/or biased objects. We obtained the striking result of reducing the intrinsic dispersion of the luminosity relation down to $\delta\approx0.12$ dex.
   For a subsample of 30 sources at redshift $z$\,$\approx$\,3 we were able to further reduce $\delta$ to $0.09$ dex (which is entirely accountable for considering quasar variability and inclination effects), thanks to the fact that in the said narrow redshift window we have high-quality pointed X-ray observations for roughly half of the subsample. This, coupled with the non-evolution of the UV SED, X-ray spectra and slope of the relation $\gamma$, with respect to low-redshift QSOs, single-handedly proves the robustness of the usage of quasars as standardizable candles: in order to invalidate the standardizability of quasars there would need to be a step evolution for the sole parameter $\beta$ in the region not covered by our data, which is highly unlikely. Furthermore, our work shows how crucial the possibility is of using high-quality X-ray observation for the determination of the unabsorbed X-ray flux of cosmological quasars.

\begin{acknowledgements}
EL acknowledges the support of grant ID: 45780 Fondazione Cassa di Risparmio Firenze.
\end{acknowledgements}

   \bibliographystyle{aa} 
   \bibliography{biblio} 

\appendix

\section{X-ray and optical/UV spectra}
\label{X-ray and optical/UV spectra}
\begin{figure*}
   \centering
   \includegraphics[width=\hsize]{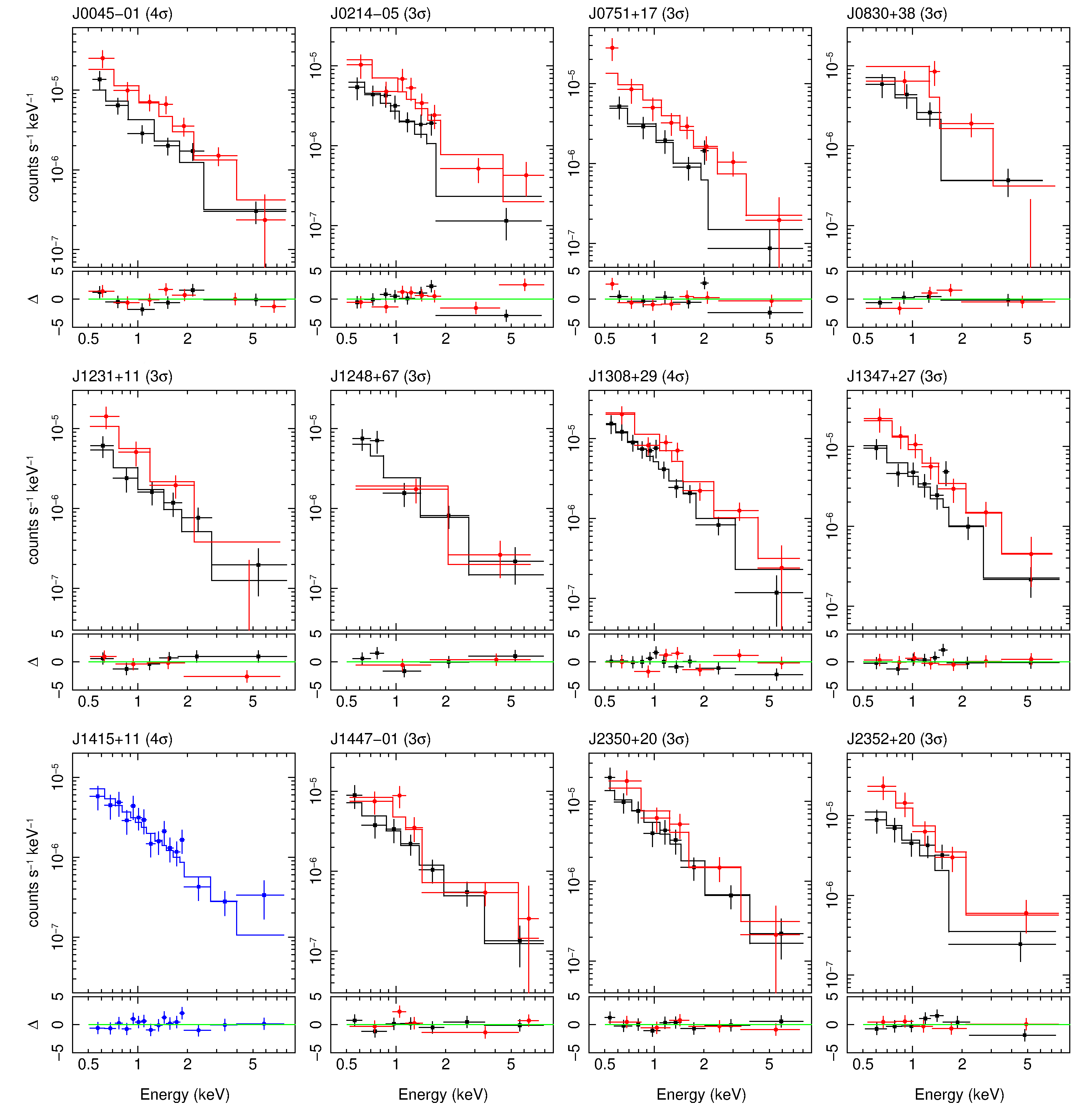}
   \caption{X-ray spectra and best fit models for the 12 sources in the $z=3.0$--3.3 sample with no previous analysis in the literature. Black and red data points indicate {\it XMM-Newton} EPIC-PN and EPIC-MOS cameras respectively, blue points indicate Chandra data. After the source name, the significance (in units of $\sigma$) adopted for the graphical rebinning of the spectra is indicated.}
              \label{fig:x_spec}%
    \end{figure*}

\begin{figure*}
   \centering
   \includegraphics[width=\hsize]{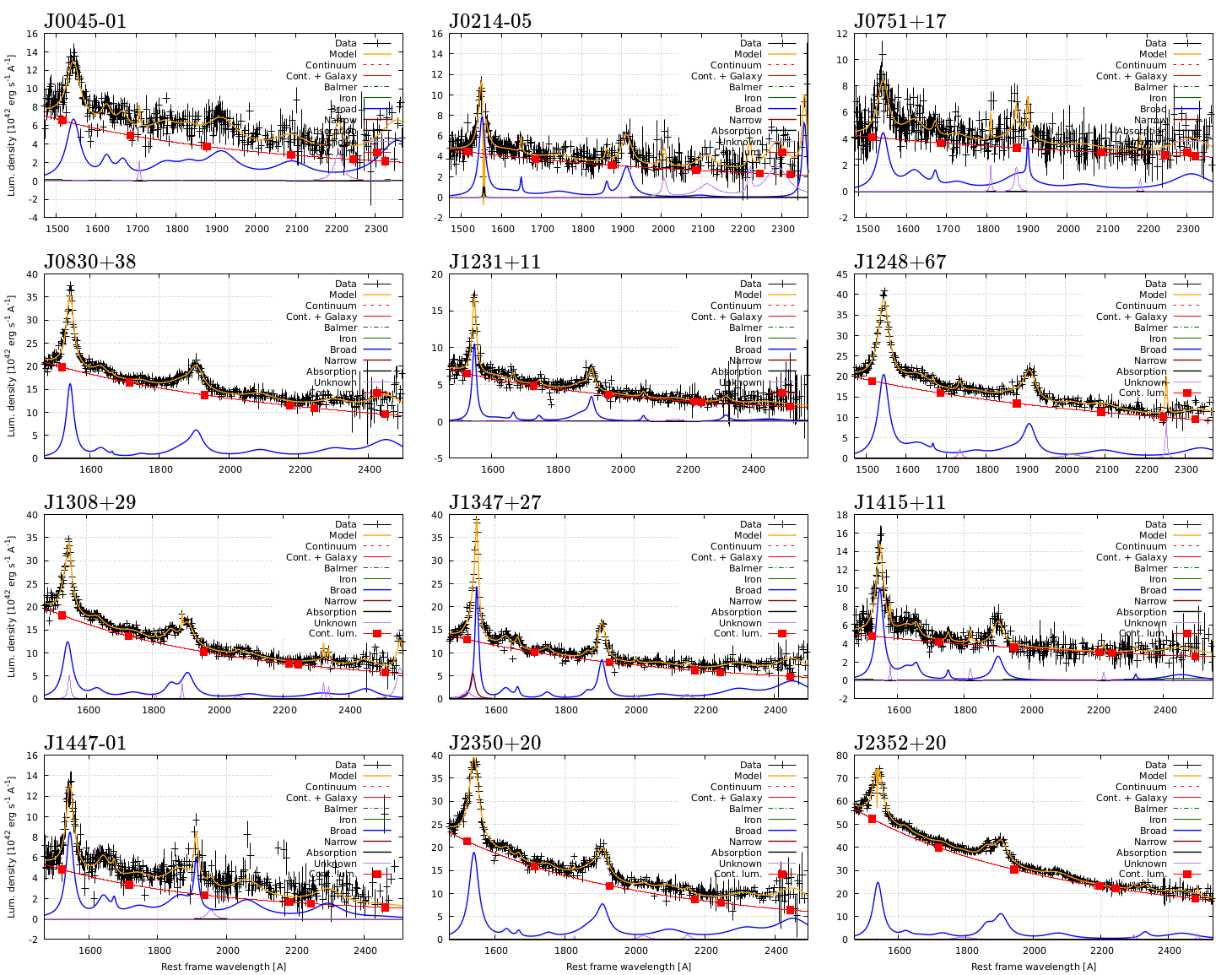}
   \caption{Same as figure \ref{fig:x_spec} for the SDSS optical (rest-frame UV) spectra. The lines listed in the legend correspond to the different models employed in the spectral fitting procedure.}
              \label{fig:specUV}%
    \end{figure*}

   We performed a complete X-ray and optical/UV spectral analysis for all the sources in our sample as described in Section 2. Fifteen of the sources of the $z$\,$\sim$\,3 ``golden'' subsample have already been extensively studied, and their X-ray and optical/UV spectra have been published in dedicated papers \citep{nar19,lus21}.\\
   Figures \ref{fig:x_spec} and \ref{fig:specUV} show the X-ray and optical/UV spectra for the 12 sources that do not have published spectra to date.

\end{document}